\documentclass{aa}
\usepackage{url}
\usepackage{graphicx}
\usepackage{natbib}
\usepackage{txfonts}
\begin{document}
   \title{X-ray spectroscopy and photometry of the long-period polar
     AI~Tri with XMM-Newton
     \thanks{Based on observations obtained with XMM-Newton, an ESA science
     mission with instruments and contributions directly funded by ESA Member
     States and NASA.}
   }

   \subtitle{}

   \titlerunning{X-ray spectroscopy and photometry of AI~Tri with
   XMM-Newton}

   \author{I.~Traulsen   \inst{1}
     \and  K.~Reinsch    \inst{1}
     \and  R.~Schwarz    \inst{2}
     \and  S.~Dreizler   \inst{1}
     \and  K.~Beuermann  \inst{1}
     \and  A.~D.~Schwope \inst{2}
     \and  V.~Burwitz    \inst{3}
          }

   \offprints{I.~Traulsen}

   \institute{Institut f\"ur Astrophysik, Georg-August-Universit\"at
              G\"ottingen, Friedrich-Hund-Platz 1, 37077 G\"ottingen,
              Germany\\
              \email{traulsen@astro.physik.uni-goettingen.de}
         \and
             Astrophysikalisches Institut Potsdam, An der Sternwarte 16, 14482
              Potsdam, Germany
	 \and
	     Max-Planck-Institut f\"ur extraterrestrische Physik, Postfach
              1312, 85741 Garching
             }

   \date{Received August 28, 2009; accepted March, 6 2010}

%
%---------- ABSTRACT ----------
%
  \abstract
%%% context heading (optional)
   {The energy balance of cataclysmic variables with strong magnetic fields is
     a central subject in understanding accretion processes on magnetic white
     dwarfs. With XMM-Newton, we perform a spectroscopic and photometric study
     of soft X-ray selected polars during their high states of accretion.}
%%% aims heading (mandatory)
   {On the basis of X-ray and optical observations of the magnetic cataclysmic
     variable AI~Tri, we derive the properties of the spectral components,
     their flux contributions, and the physical structure of the accretion
     region in soft polars.}
%%% methods heading (mandatory)
   {We use multi-temperature approaches in our \textsc{xspec} modeling of the
     X-ray spectra to describe the physical conditions and the structures of
     the post-shock accretion flow and the accretion spot on the white-dwarf
     surface. In addition, we investigate the accretion geometry of the system
     by a timing analysis of the photometric data.}
%%% results heading (mandatory)
    {Flaring soft X-ray emission from the heated surface of the white dwarf
      dominates the X-ray flux during roughly 70\% of the binary cycle.  This
      component deviates from a single black body and can be described by a
      superimposition of mildly absorbed black bodies with a Gaussian
      temperature distribution between $kT_\mathrm{bb,low} := 2\,\mathrm{eV}$
      and $kT_\mathrm{bb,high} = 43.9^{+3.3}_{-3.2}\,\mathrm{eV}$, and
      $N_\mathrm{H,ISM} = 1.5^{+0.8}_{-0.7}\times10^{20}\,\mathrm{cm}^{-2}$.
      In addition, weaker hard X-ray emission is visible nearly all the
      time. The spectrum from the cooling post-shock accretion flow is most
      closely fitted by a combination of thermal plasma \textsc{mekal} models
      with temperature profiles adapted from prior stationary two-fluid
      hydrodynamic calculations. The resulting plasma temperatures lie between
      $kT_{\mathsc{mekal},\mathrm{low}} = 0.8^{+0.4}_{-0.2}\,\mathrm{keV}$ and
      $kT_{\mathsc{mekal},\mathrm{high}} = 20.0^{+9.9}_{-6.1}\,\mathrm{keV}$;
      additional intrinsic, partial-covering absorption is on the order of
      $N_\mathrm{H,int} = 3.3^{+2.5}_{-1.2}\times 10^{23}\,\mathrm{cm}^{-2}$.
      The soft X-ray light curves show a dip during the bright phase, which
      can be interpreted as self-absorption in the accretion
      stream. Phase-resolved spectral modeling supports the picture of
      one-pole accretion and self-eclipse. One of the optical light curves
      corresponds to an irregular mode of accretion. During a short XMM-Newton
      observation at the same epoch, the X-ray emission of the system is
      clearly dominated by the soft component.}
%%% conclusions heading (optional)
   {}

%
%---------- KEYWORDS ----------
%

   \maketitle

%
%---------- INTRODUCTION ----------
%
\section{Introduction}

  \object{AI~Tri} (RX\,J0203.8+2959) was first described within a sample of
  ROSAT-discovered bright soft X-ray sources by \citet{beuermann:93}. Their
  classification of AI~Tri as an AM~Her type binary (also called a polar) was
  later confirmed in a multiwavelength study by \citet{schwarz:98}, who
  identified cyclotron humps in optical spectra obtained during a high state
  of accretion. The orbital period of $P_\mathrm{orb} = 4.6\,\mathrm{hrs}$ is
  one of the longest known among polars, whereas the magnetic field strength
  of $B = 38\pm2\,\mathrm{MG}$ and the amplitude of the long-term brightness
  variations between $V = 18\fm0 - 15\fm5$ \citep{schwarz:98} lie in the
  typical parameter range of this class.  Based on the wavelength dependence
  of the $UBVRI$ light curve minima and the variations in both the linear and
  the circular polarization, \citet{katajainen:01} suggest that the system has
  a high inclination of $i\approx 70\degr\pm 20\degr$ and accretes onto two
  almost equally fed magnetic poles.  On the other hand, \citet{schwarz:98}
  propose that a single dominating accretion region is active at the epoch of
  their observations.

  AI~Tri belongs to a significantly large group of AM~Her systems that were
  found to emit almost entirely at X-ray energies below 0.5\,keV during the
  ROSAT All-Sky Survey \citep{beuermann:93,thomas:98,beuermann:99}. Although
  these systems could play an important role in investigating the energy
  balance of polars, only a few of them have been studied using
  high-resolution X-ray spectroscopy \citep{ramsay:03,ramsay:04shortper}. We,
  therefore, initiated dedicated observations with XMM-Newton to perform a
  detailed study of the spectral components, their flux contributions, and the
  physical structure of the accretion region of polars selected by their
  distinct soft X-ray fluxes during high-states of accretion.  In the
  following, we present an analysis of the magnetic cataclysmic variable
  AI~Tri based on new XMM-Newton and optical data, and archival ROSAT data.

%% Observation log
  \begin{table*}
    \caption{\footnotesize{Log of the ROSAT and XMM-Newton observations and of
        the optical photometry of AI~Tri.}}
    \label{tab:obslog}      
    {\centering
    \begin{tabular}{c@{\qquad}l@{\qquad\quad}c@{\qquad\quad}c@{\qquad}r
	@{\qquad}r@{\qquad}c@{\qquad}l}
      \hline\hline       
      Date & Telescope\tablefootmark{a} & Instrument & Filter &
      $t_\mathrm{exp} [\mathrm{s}]$ & $t_\mathrm{cycle} [\mathrm{s}]$ &
      Duration [h] & Observer\\ 
      \hline                    
      1998 Jan 15$-$Feb 05 & ROSAT & HRI & 0.1$-$2.4\,keV & & & 9.6
         & PI Schwarz\\
      2005 Aug 15 & XMM-Newton & EPIC/pn & 0.1$-$10\,keV & & & 0.3
         & PI Reinsch\\
      2005 Aug 15 & XMM-Newton & EPIC/MOS & 0.1$-$10\,keV & & & 1.4
         & PI Reinsch\\
      2005 Aug 15 & XMM-Newton & OM & UVM2 & & & 1.1 & PI Reinsch\\
      2005 Aug 15 & XMM-Newton & RGS & 0.3$-$2.5\,keV  & & & 1.3 & PI Reinsch\\
      2005 Aug 22 & XMM-Newton & EPIC & 0.1$-$10\,keV & & & 5.6 & PI Reinsch\\
      2005 Aug 22 & XMM-Newton & OM & UVM2 & & & 5.6 & PI Reinsch\\
      2005 Aug 17 & AIP 70\,cm & TK1024-01 & V & 60 & 66 & 6.5 & Schwarz\\
      2005 Aug 29 & AIP 70\,cm & TK1024-01 & V & 60 & 66 & 7.6 & Schwarz\\
      2006 Nov 09 & IAG 50\,cm & STL-6303E & WL & 180 & 191 & 6.6 & Traulsen\\
      2006 Nov 15 & IAG 50\,cm & STL-6303E & V & 240 & 251 & 4.7 & Traulsen\\
      2006 Nov 16 & IAG 50\,cm & STL-6303E & V & 240 & 247 & 1.8 & Traulsen\\
      2007 Jan 14 & IAG 50\,cm & STL-6303E & V &  90$-$180 &  97$-$187 & 2.2
         & Traulsen\\
      2007 Jan 25 & AIP 70\,cm & TK1024-01 & V &  90 & 95 & 4.6 & Schwarz\\
      2007 Jan 29 & \textsc{Monet}/North & Alta E47+ & V &  15 &  18 & 1.8 &
           Hessman\\
      2007 Jan 30 & \textsc{Monet}/North & Alta E47+ & V &  15 &  18 & 2.0 &
           Hessman\\
      2007 Mar 14 & IAG 50\,cm & STL-6303E & V & 150 & 157 & 1.8 & Traulsen\\
      2007 Mar 14 & AIT 80\,cm & STL-1001E & V &  20$-$30 &  23$-$33 & 1.5
         & Nagel\\
      \hline                  
    \end{tabular}
    }
    \tablefoottext{a}{AIP: Astrophysikalisches Institut Potsdam. IAG:
      Institut f\"ur Astrophysik G\"ottingen. AIT: Institut f\"ur Astronomie
      und Astrophysik T\"ubingen, Abt.~Astronomie (Kepler Center for Astro and
      Particle Physics).}
  \end{table*}

%
%---------- OBSERVATIONS ----------
%
\section{Observations and data reduction}

\subsection{X-ray observations}

  With XMM-Newton, we obtained a 20\,ksec (5.6\,hrs) exposure during a high
  state of AI~Tri on August 22, 2005 (observation ID 0306841001), which covers
  for the first time uninterruptedly more than one complete binary orbit. The
  EPIC instruments were operated in full frame mode with the thin filter.  The
  UV light curve at an effective wavelength of 2310\,{\AA} was measured
  simultaneously using the Optical Monitor (OM) in fast mode with the UVM2
  filter. RGS spectra were not used from this pointing, since the X-ray flux
  of the object was too low, and the net source count rate measured by the RGS
  instruments was consistent with zero for more than 80\,\% of the exposure
  time.

\begin{figure}
    \centering
    \includegraphics[width=8.7cm]{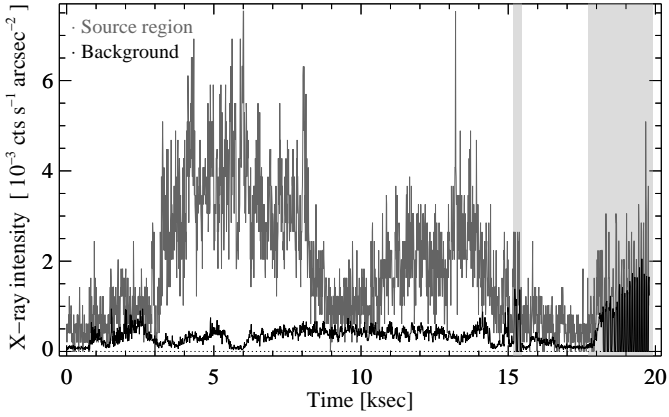}
    \caption{\footnotesize{EPIC/pn X-ray intensity in both the source region
        (source plus background) and the source-free background region during
        the 20\,ksec XMM-Newton observation, binned into time intervals of
        10\,s.}}
    \label{fig:bkg}%
  \end{figure}

  Data were reduced using the XMM-Newton Software Analysis System \textsc{sas}
  v8.0. Large parts of the exposure were affected by an enhanced background
  signal.  We omitted two time intervals from the analysis
  (Fig.~\ref{fig:bkg}): a strong soft proton flare after 15\,ksec and the last
  2\,ksec of the observation, when the satellite approaches the Van Allen
  radiation belt.  During these intervals, a considerable spectral hardening
  at energies above 3\,keV and spectral softening at energies below 1\,keV
  were detected.

  When extracting light curves, spectra, and images, single and double events
  were processed in the case of EPIC/pn data, supplemented by quadruple events
  in EPIC/MOS data. Phase-resolved spectra were generated during the four
  phase intervals defined in Sect.~\ref{sec:photo}. Source data were collected
  from an aperture of 25\,arcsec around the source center, background data
  from a larger source free region close to the source position.  To diminish
  potential pile-up effects, we excluded the innermost 5\,arcsec of the source
  region when extracting the mean EPIC/pn spectrum and the bright-phase
  spectra, where the peak count rates (source plus background) reach
  $6-8\,\mathrm{cts\,s}^{-1}$ for EPIC/pn and $3-4\,\mathrm{cts\,s}^{-1}$ for
  EPIC/MOS, respectively. Appropriate spectral response matrices were
  generated for the individual instruments and extraction regions applying
  calibration files from the July 2008 release.  Spectral bins comprise a
  minimum of 20~cnts (mean EPIC/pn spectrum: 30~cnts).  Photon arrival times
  were corrected to the barycenter of the solar system via the
  \textsc{barycen} task.

  In the soft energy regime below 0.9\,keV, where the calibration is known to
  be least precise \citep{crosscalib,mateos:09}, we use the isolated neutron
  star \object{RX\,J1856.4$-$3754} to calibrate the EPIC
  spectra. RX\,J1856.4$-$3754 was established as a low-energy calibration
  target for missions such as Chandra, Suzaku, Swift, and
  XMM-Newton,\footnote{International Astronomical Consortium for High Energy
    Calibration, \url{http://www.iachec.org}} and exhibits a high soft X-ray
  and low optical flux.  Its spectrum is free from prominent line features and
  described well by two black bodies \citep{beuermann:06}. We compare archival
  XMM-Newton EPIC data of RX\,J1856.4$-$3754, which span more than five years,
  with the model spectra of \citet{beuermann:06} and carefully adjust the
  instrumental effective areas. In this approach, we assume that the energy
  redistribution is independent of the incident energy when considering only
  objects with very soft X-ray spectra. The discrepancies between data and
  model remain mostly below $5\,\%$, with larger excesses up to $20\,\%$ for
  EPIC/MOS2.

  A first XMM-Newton observation of AI~Tri on August 15, 2005 in the same
  configuration as described above resulted in partially corrupted data
  (observation ID 0306840901) due to technical problems.  During 1.1 to
  5\,ksec (0.3 to 1.4\,hrs), EPIC, RGS (1st order), and OM data are usable. We
  apply the same data reduction steps as applied to the data acquired on
  August 22, 2005.  With peak count rates of $21\,\mathrm{cts\,s}^{-1}$
  collected by EPIC/pn and $1.4\,\mathrm{cts\,s}^{-1}$ by the RGS detectors,
  the source flux was considerably higher than one week later.  Therefore, we
  extended the EPIC/pn aperture to 30\,arcsec, excising the innermost
  7\,arcsec. The EPIC spectra were rebinned to at least 20 counts per bin.

  In addition, we used unpublished ROSAT archival data obtained between
  January 15 and February 5, 1998 with the HRI detector. The total integration
  time of 35\,ksec (9.6\,hrs) consisted of nine observation intervals. The
  individual light curves were extracted with standard tasks of the
  \textsc{exsas/Midas} \citep{zimmermann:93} software packages, corrected to
  the barycenter, and combined to a common light curve profile.

%% Background

%----------------------------------
\subsection{Optical data}

  At the G\"ottingen 50\,cm, the Potsdam 70\,cm, the \textsc{Monet}/North
  (Texas) 1.2\,m, and the T\"ubingen 80\,cm telescopes, we carried out optical
  $V\!$-band and white-light photometry during ten nights between August 2005
  and March 2007. AI~Tri was found in high or intermediate high states at
  estimated brightnesses between $V = 17\fm0$ and $15\fm5$.  For data
  reduction and differential photometry, we used \textsc{tripp}
  \citep{schuh:03} and \textsc{ESO-Midas} routines. Light curves were
  determined using differential aperture photometry.  Apparent magnitudes of
  the system were calculated against the $V\!$-band magnitudes of the
  comparison stars USNO-B1.0~1199$-$0026710 ($V = 13\fm29\pm0\fm05$) and
  USNO-B1.0~1199$-$0026672 ($V = 14\fm62\pm0\fm05$) determined by
  \citet{schwarz:96}. Fluxes are given relative to their orbital mean, and
  flux errors were derived by means of photon statistics. The times were
  converted to Terrestrial Time and corrected to the barycenter on the basis
  of the JPL ephemeris \citep{JPL}.  An overview of all the data obtained is
  given in Table~\ref{tab:obslog}.

%
%---------- PHOTOMETRY ----------
%
\section{Optical, UV, and X-ray light curves}
\label{sec:photo}

%----------------------------------
\subsection{Optical photometry}
\label{sec:optphot}

\begin{figure}
    \centering
    \includegraphics[width=8.7cm]{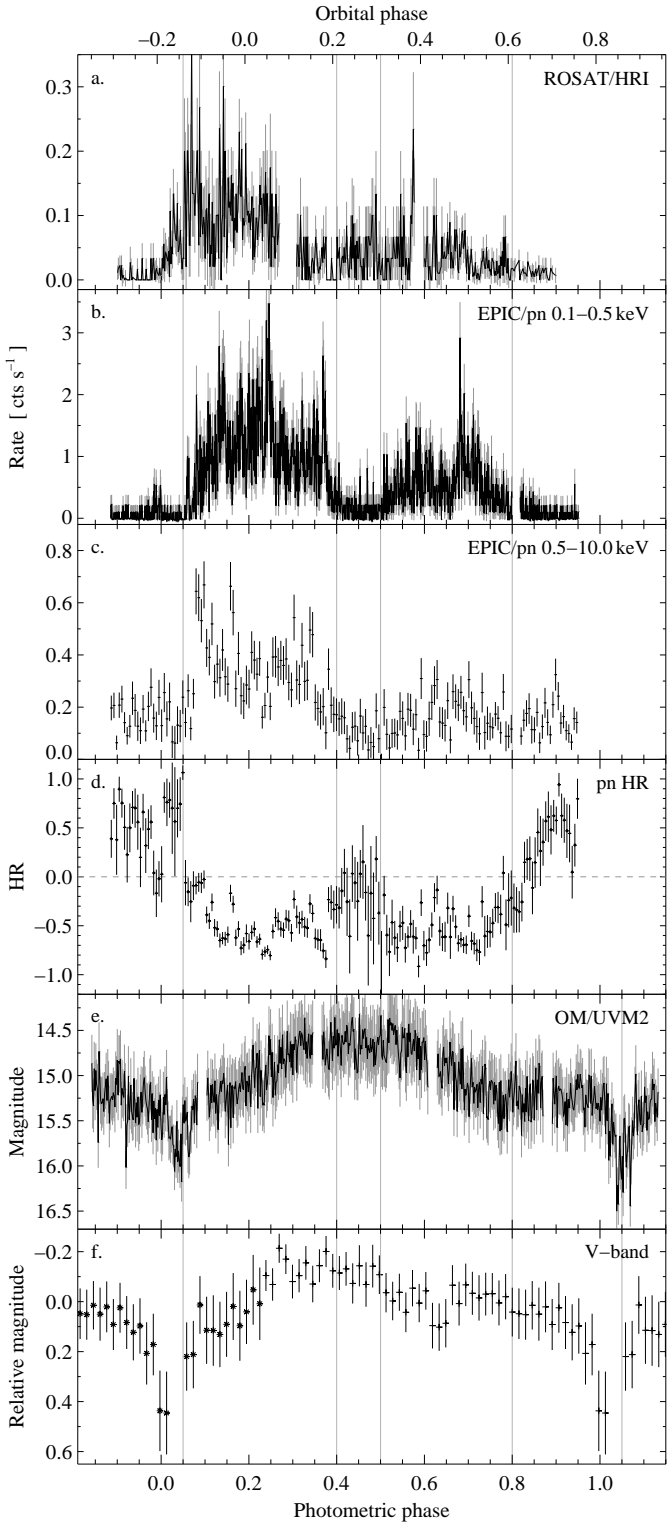}
    \caption{\footnotesize{X-ray, ultraviolet, and optical light curves of
        AI~Tri folded over photometric phase. Orbital phase as
        $\varphi_\mathrm{orb}=\varphi_\mathrm{phot}-0.191$ according to
        \citet{schwarz:98} is marked at the top of the plot. \textit{a.}
        ROSAT/HRI data of Jan./Feb.~1998 with a total integration time of
        35\,ksec, combined to a common light curve profile and rebinned into
        30\,s intervals.  \textit{b.-d.} EPIC/pn light curve on August 22,
        2005, divided into the soft component S, the hard component H, and the
        associated hardness ratio HR$=$(H$-$S)/(H$+$S). The soft band data
        have been binned into 10\,s, the hard band and the hardness ratio into
        100\,s.  \textit{e.} OM light curve in the ultraviolet UVM2 filter at
        an effective wavelength of 2310\,{\AA} in time bins of 30\,s.
        \textit{f.}  Optical $V\!$-band data obtained at the G\"ottingen
        50\,cm telescope on November 15, 2006, plotted twice for clarity.}}
    \label{fig:lc1044}%
  \end{figure}

  The optical $V\!$-band light curves of AI~Tri (e.\,g.
  Fig.~\ref{fig:lc1044}f) during high and intermediate high states of
  accretion are largely similar to the data presented by \citet{katajainen:01}
  and \citet{schwarz:98}. The orbital modulation deviates from a sinusoidal
  shape and is slightly asymmetric showing a steeper rise and smoother
  decline. An irregular pattern of narrow small dips is superimposed on the
  light curves, one dip recurring at photometric phase zero, which has not
  been seen in earlier data.

  The orbital period of AI~Tri has not been known accurately enough to
  extrapolate the photometric ephemeris given by \citet{schwarz:98} to the
  epochs of our observations. \citet{schwarz:98} established photometric phase
  zero at the broad minima of the optical high-state light curves, the only
  recurrent feature that could be identified in the vast majority of the light
  curves. Using timings of six new $V\!$-band minima mapped between August
  2005 and March 2007 in addition to the data published by \citet{schwarz:98},
  we updated their photometric ephemeris. We determined the minima by
  performing Gaussian fits to the light curves, excluding the dips
  (Table~\ref{tab:minima}). By means of a least-squares method, the inverse
  square sum of observed minus calculated minimum times $(O-C)^{-2}$ was
  minimized for narrowly spaced trial periods within a $\pm10\,\sigma$ range
  of the \citet{schwarz:98} value. We derived the improved $V\!$-band
  ephemeris
  \begin{center}
    $\mathrm{BJD}_\mathrm{min}(TT) = 2\,451\,439\fd0391(10) + 0\fd19174566(9)
    \times E$,
  \end{center}
  which we use throughout this paper.

  In addition, \citet{schwarz:98} determined an orbital ephemeris of the
  system from long-slit spectroscopy. They defined $\varphi_\mathrm{orb} = 0$
  at the blue-to-red zero crossing of the radial velocities, which corresponds
  to $\varphi_\mathrm{phot} = 0.191 \pm 0.080$, and identified it with the
  inferior conjunction of the secondary. A constant relation between
  photometric and orbital phase can be assumed, if the system rotates
  synchronously, as most polars do. \citet{schwarz:98} derived a degree of
  synchronism better than $|P_\mathrm{orb}-P_\mathrm{spin}|/P_\mathrm{orb} <
  10^{-4}$ for AI~Tri. We show the orbital phasing as
  $\varphi_\mathrm{phot}-0.191$ at the top of Fig.~\ref{fig:lc1044}, since it
  can provide geometrical information about the system
  (Sect.~\ref{sec:geometry}).

%% Minima
  \begin{table}
    \caption{\footnotesize{Barycentric timings of the $V\!$-band minima of
        AI~Tri.}}
    \label{tab:minima}
    {\centering
    \begin{tabular}{l@{\qquad}c@{\qquad}r@{\qquad}r}
      \hline\hline
      \hspace*{2.ex}$\mathrm{BJD}_\mathrm{min}$(TT) & $\Delta T_\mathrm{min}$
      & $O-C$~ & Cycle~ \\ \hline
   $2\,449\,243.93702$\tablefootmark{a}  &  0.0008  &    0.0114 & $-$11\,448 \\
   $2\,450\,042.36390$\tablefootmark{a}  &  0.0017  &    0.0008 &  $-$7\,284 \\
   $2\,450\,043.51363$\tablefootmark{a}  &  0.0009  & $-$0.0031 &  $-$7\,278 \\
   $2\,450\,046.39803$\tablefootmark{a}  &  0.0029  &    0.0398 &  $-$7\,263 \\
   $2\,450\,047.54570$\tablefootmark{a}  &  0.0014  &    0.0252 &  $-$7\,257 \\
   $2\,450\,049.26491$\tablefootmark{a}  &  0.0013  & $-$0.0087 &  $-$7\,248 \\
   $2\,450\,049.45553$\tablefootmark{a}  &  0.0011  & $-$0.0146 &  $-$7\,247 \\
   $2\,450\,122.32708$\tablefootmark{a}  &  0.0019  &    0.0281 &  $-$6\,867 \\
   $2\,450\,123.28113$\tablefootmark{a}  &  0.0013  &    0.0037 &  $-$6\,862 \\
   $2\,450\,423.36107$\tablefootmark{a}  &  0.0009  & $-$0.0068 &  $-$5\,297 \\
   $2\,450\,748.74298$\tablefootmark{a}  &  0.0047  & $-$0.0613 &  $-$3\,600 \\
   $2\,450\,748.95113$\tablefootmark{a}  &  0.0063  &    0.0242 &  $-$3\,599 \\
   $2\,453\,612.47355$                   &  0.0003  & $-$0.0133 &    11\,335 \\
   $2\,454\,055.59847$                   &  0.0014  & $-$0.0096 &    13\,646 \\
   $2\,454\,115.23645$                   &  0.0006  &    0.0169 &    13\,957 \\
   $2\,454\,126.35745$                   &  0.0007  &    0.0156 &    14\,015 \\
   $2\,454\,129.62165$                   &  0.0008  &    0.0392 &    14\,032 \\
   $2\,454\,174.27428$                   &  0.0072  & $-$0.0865 &    14\,265 \\
      \hline
    \end{tabular}
    }
    \tablefoottext{a}{\citet{schwarz:98}}
  \end{table}

%% Multiwavelength light curves regular

%% Multiwavelength light curves irregular
  \begin{figure}
    \centering
    \includegraphics[width=8.7cm]{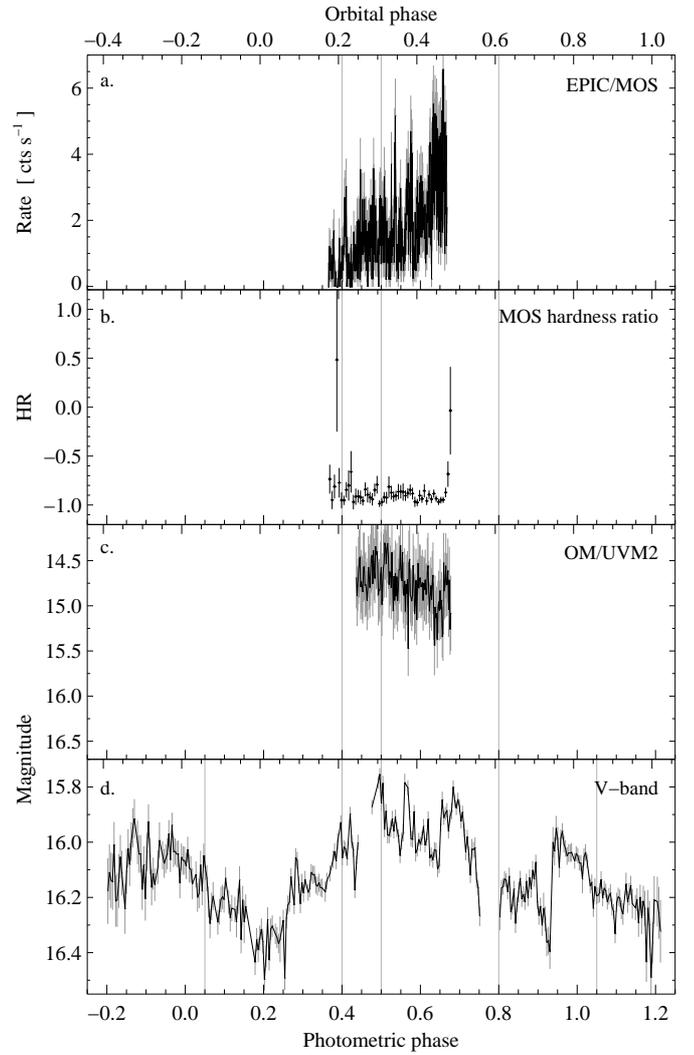}
    \caption{\footnotesize{Multi-wave-band light curves obtained with
        XMM-Newton on Aug.~15 \textit{(a.-c.)} and with the Potsdam 70\,cm
        telescope on Aug.~17, 2005 \textit{(d.)}. The X-ray data are binned
        into 10\,s intervals, the ultraviolet into 30\,s, and the hardness
        ratios into 100\,s, respectively.  Grey lines indicate the phases of
        high and low soft X-ray flux in the 20\,ksec exposure.}}
    \label{fig:lc1041}%
  \end{figure}

%----------------------------------
\subsection{XMM-Newton light curves}
\label{sec:xmmphot}

  The observation with the Optical Monitor and the UVM2 filter at an effective
  wavelength of 2310\,{\AA} presents the first mid ultraviolet light curve of
  AI~Tri (Fig.~\ref{fig:lc1044}e). It exhibits a roughly similar shape but
  larger amplitude than the optical light curves, an almost sinusoidal
  variation, and an additional dip in the light curve minimum near
  $\varphi_\mathrm{phot} = 0.05$. This dip is shifted by about $\Delta\varphi
  = 0.05$ in phase relative to the dip within the optical minimum. The UV
  maximum at $\varphi_\mathrm{phot} = 0.45$ appears to be flattened, but due
  to the lack of additional UV data it remains unclear whether this feature
  recurs or is unique.

  From the 20\,ksec exposure performed by the EPIC instruments, we extracted
  the first continuous X-ray light curves of AI~Tri over a full binary orbit
  (Fig.~\ref{fig:lc1044}b$-$c). Two X-ray bright phases during the intervals
  $\varphi_\mathrm{phot} = 0.05-0.40$ and $\varphi_\mathrm{phot} =$ $0.50$ $-$
  $0.80$, as defined by the ephemeris given above, commence with a steep rise
  at $\varphi_\mathrm{phot} = 0.05$. They are interrupted by a short soft
  minimum at $\varphi_\mathrm{phot} = 0.40-0.50$ and a larger interval of low
  flux between $\varphi_\mathrm{phot} = 0.80$ and 1.05 (faint phase). The flux
  variation is most pronounced in the soft energy band, the bright phases
  lasting about 70\,\% of the orbital period, while hard X-ray emission
  remains visible at a low level at all times. The hard X-ray flux increases
  by a factor of about three at the beginning of the first bright soft phase
  and varies slightly during the second one.

  The hardness ratio HR$=$(H$-$S)/(H$+$S), where H and S represent the counts
  at energies above and below 0.5\,keV, respectively, correlates largely with
  the light curves in the soft energy band, and tends to $+1$ during faint and
  to $-1$ during bright phases (Fig.~\ref{fig:lc1044}d). Throughout the soft
  minimum at $\varphi_\mathrm{phot} = 0.45$, the hardness ratio levels off
  around 0 and thus differs from the faint-phase behavior, where it rises to
  values around 0.7. Similar characteristics of the light curves are found
  from the ROSAT/HRI data between January 15 and February 5, 1998
  (Fig.~\ref{fig:lc1044}a).

%----------------------------------
\subsection{Irregular mode on August 17, 2005}
\label{sec:irreg}

  The $V\!$-band light curve on August 17, 2005 (Fig.~\ref{fig:lc1041}d)
  differs markedly from the majority.  At a magnitude around $V = 16\fm0$
  (close to that during high states of accretion), the light curve is
  strikingly variable and more asymmetric than in normal high states; the
  photometric minimum appears to be shifted by 0.2 in phase. In consequence,
  we did not include this peculiar light curve in the determination of the new
  ephemeris.  During the next observation on August 29, the $V\!$-band light
  curve fits in the prevailing shape again. \citet{schwarz:98} described a
  comparable phase-shift with an altered light curve shape in their $RV$
  photometry of October and November 1992 as an `irregular mode'.

  The irregular light curve was observed only two days after the curtailed
  XMM-Newton exposure on August 15, 2005 (Fig.~\ref{fig:lc1041}a-c).  The
  EPIC/MOS and RGS1 data cover phases $\varphi_\mathrm{phot} = 0.35-0.65$ and
  EPIC/pn $\varphi_\mathrm{phot} = 0.58-0.65$, respectively.  This is
  equivalent to the soft minimum phase and about the first half of the second
  bright phase of the August 22 light curves. The EPIC count rate in the soft
  energy band, marked by pronounced flickering, is about a factor of eight
  higher than one week later. In contrast, little hard X-ray emission was
  registered, not exceeding the low level of hard X-rays on August 22. The
  hardness ratios remain almost constant at about $-0.87$, even during the
  $\varphi_\mathrm{phot} = 0.4-0.5$
  interval, where they increase to zero in the August 22 data. The ultraviolet
  light curve has similar magnitudes as on August 22. Owing to the
  insufficient coverage of the orbital cycle, the appearance of the soft X-ray
  minimum and a potential phase shift cannot be determined.

%----------------------------------
\subsection{Short-period variations}
\label{sec:shortperiod}

\begin{figure}
    \centering \includegraphics[width=8.7cm]{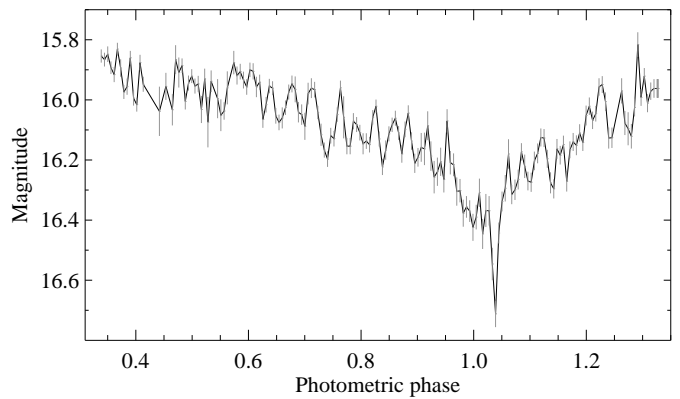}
    \caption{\footnotesize{$V\!$-band light curve acquired on January 25, 2007
        at the 70\,cm telescope of the Astrophysical Institute Potsdam.  Cycle
        time is 100\,s.}}
    \label{fig:lc_aip}%
  \end{figure}

\begin{figure}
    \centering
    \includegraphics[width=8.7cm]{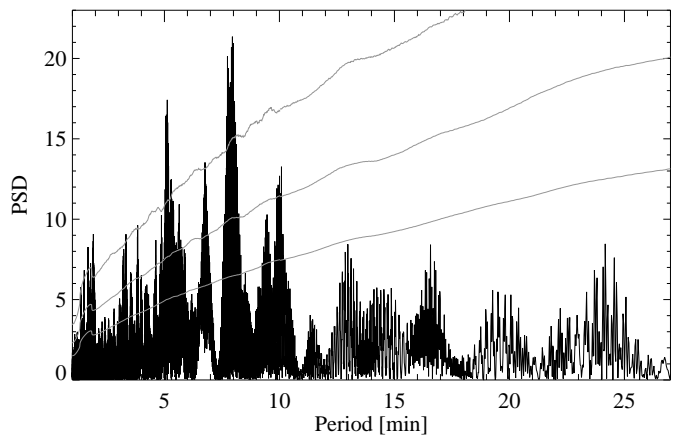}
    \caption{\footnotesize{Periodogram based on the highest time resolved
        $V\!$-band data from Jan.~2007. In grey, the 95\%, 99\%, and 99.9\%
        significance levels.}}
    \label{fig:period}%
  \end{figure}

  The multiwavelength light curves are highly variable and exhibit, on
  timescales as short as one minute, strong and potentially recurring
  flickering, which is typical of soft polars. \citet{schwarz:98} describe
  possible quasi-periodic oscillations (QPOs) at $6.5-7$\,min and
  $13.5-14$\,min in their optical data. We examine three highly resolved light
  curves with cycle times below 100\,s, obtained between January 25 and 30,
  2007, to investigate this feature.  Substructure on a timescale of several
  minutes is clearly identifiable, as demonstrated for instance in
  Fig.~\ref{fig:lc_aip}. After subtracting a strongly smoothed light curve to
  remove the low-frequency variations, a Lomb-Scargle analysis
  \citep{lomb:76,scargle:82} was performed to search for periodicity. The
  periodogram in Fig.~\ref{fig:period} shows the typical broad maxima, which
  could be associated with quasi-periodic flickering, but may also arise from
  a superimposition of different patterns. To exclude false positives, the
  influence of aperiodic, random brightness variations has to be considered.
  Such low-frequency flickering or 'red noise' is caused by unsteady mass
  transfer and has to be distinguished from the more regular, quasi-periodic
  oscillations, excited by larger blobs in the accretion stream. To reproduce
  the red-noise and window effects in the observations, the observed power
  spectral density (PSD) is fitted with a power law over frequency, and light
  curves with similar mean count rate, standard deviation, and PSD shape are
  simulated on the time grid of the observed data.  We adopt the Monte Carlo
  approach of \citet{benlloch:01}, which they use in the search for QPOs from
  active galactic nuclei, and determine confidence levels from the
  periodograms of 10\,000 fake light curves with the same power-law-shaped
  PSD. The corresponding levels of 99.9\,\%, 99\,\%, and 95\,\% significance
  are overplotted in Fig.~\ref{fig:period}. At two periods around 8\,min and
  5\,min, the power spectrum exceeds 99.9\% significance, so the features may
  be considered to be real. The QPO frequencies found by \citet{schwarz:98}
  are not significant at the 99.9\% level.

  No low-frequency periodicity was found in the XMM-Newton ultraviolet and
  soft X-ray light curves.  All spikes in the periodogram lie well below the
  90\,\% confidence level.  Rapid variations in the soft X-ray light curves
  are obvious, but no regular or periodic pattern is detectable.

%% High-res optical

% Periodogram QPOs

%
%---------- SPECTROSCOPY ----------
%
\section{X-ray spectroscopy}
\label{sec:spectra}

  With XMM-Newton, we obtained the first well resolved X-ray spectra with full
  orbital phase coverage of AI Tri. We used them to constrain the physical
  conditions and the structure of the accretion region on the white dwarf
  surface and of the post-shock accretion flow, and to compare the overall
  flux contributions from both origins. For this purpose, we analyzed the
  background-subtracted X-ray spectra by means of multi-component models using
  \textsc{xspec} v12.5 \citep{arnaud:96,dorman:03}. Initially, we modeled the
  time-averaged spectrum with single-temperature approaches for the soft and
  hard X-ray components. More complex and physically more relevant models were
  considered to improve the quality and the validity of the fits. In this
  context, we investigated separate fits to the spectra in the energy bands
  above and below 0.5\,keV, and combined fits to the total spectrum. In a
  second step, we analyzed the average spectra of four characteristic phase
  intervals to constrain the accretion geometry. In all fits, we described the
  interstellar absorption using the \textsc{tbnew} model\footnote{Based on
    \textsc{tbvarabs} in \textsc{xspec} \citep{wilms:00}, \textsc{tbnew}
    provides refined cross-sections. See\\
    \url{http://pulsar.sternwarte.uni-erlangen.de/wilms/research/tbabs/}.}
  with the abundances of \citet{wilms:00} and the cross-sections of
  \citet{verner:96a} and \citet{verner:96b}. EPIC/pn and MOS data are modeled
  simultaneously. Treating the different instruments individually does not
  improve the fit quality, as tests via F-statistics confirmed. All errors are
  given at a 90\,\% level of confidence in the following.

%% Fit results
  \begin{table*}
    \caption{\footnotesize{Results of the best \textsc{xspec} fits to the
        total and to the phase-resolved EPIC spectra of AI~Tri.}}
    \label{tab:xspecfits} 
    {\centering          
    \begin{tabular}{l@{~~~}*{4}{c@{~~}}ccr@{}lc@{~~}r@{}l@{~}c@{}r@{}l}
      \hline\hline
      Model\tablefootmark{a} & $\chi^2_\mathrm{red}$ & 
      $N_\mathrm{H,ISM}$ &
      $kT_\mathrm{bb}$ &
      $N_\mathrm{H,int}$ & cover. &
      $kT_\mathrm{M,low}$ &
      \multicolumn{2}{c}{$kT_\mathrm{M,high}$} & abund. &
      \multicolumn{2}{c}{$A_\mathrm{bb}$} &
      $L_\mathrm{bb,bol}$\tablefootmark{b} &
      \multicolumn{2}{c}{{$F_\mathrm{XMM}~~[10^{-12}$}}
      \\
      & & 
      [$10^{20}\,\mathrm{cm}^{-2}$] & 
      [eV] & 
      [$10^{23}\,\mathrm{cm}^{-2}$] &  &
      [keV] &
      \multicolumn{2}{c}{[keV]} &  &
      \multicolumn{2}{c}{[$10^6\,\mathrm{km}^2$]} &
      [$10^{32}\,\mathrm{erg\,s}^{-1}$] &
      \multicolumn{2}{c}{$\mathrm{erg\,cm}^{-2}\,\mathrm{s}^{-1}$]}
      \\
      \hline
      \multicolumn{2}{l}{\emph{2005/08/22}} & & & & & & & & & & \\
      \textsc{mekal} & $0.96$ & $1.7^{+0.8}_{-0.7}$ & $43.0^{+3.2}_{-3.0}$ &
      $3.7^{+2.4}_{-1.2}$ & $0.76^{+0.08}_{-0.12}$ & $-$ &
      $6.$&$4^{+2.6}_{-1.8}$ & $0.8^{+0.7}_{-0.4}$
      & \hspace*{1.3em}$1.$&$9^{+2.5}_{-1.0}$ & $2.2^{+1.6}_{-0.9}$
      & \hspace*{2.2em}$7.$&$2^{+3.5}_{-2.1}$ \\[0.65ex]
      2\textsc{mekal} & $0.95$ & $1.5^{+0.9}_{-0.7}$ & $43.7^{+3.4}_{-3.0}$ &
      $3.4^{+1.7}_{-1.1}$ & $0.75^{+0.05}_{-0.07}$ & $2.1^{+2.1}_{-1.0}$ &
      $9.$&$1^{+70}_{-3.0}$ & $1.0^{+1.0}_{-0.5}$ & $1.$&$6^{+1.1}_{-0.8}$ &
      $2.1^{+1.5}_{-0.8}$ & $7.$&$0^{+3.1}_{-2.0}$
      \\[0.65ex]
      3\textsc{mekal} & $0.94$ & $1.6^{+0.9}_{-0.6}$ & $44.4^{+4.3}_{-5.8}$ &
      $3.3^{+2.1}_{-1.2}$ & $0.74^{+0.08}_{-0.06}$ & $0.06^{+0.03}_{-0.02}$ &
       & & & & & & \\[0.65ex]
       & & & & & & $2.0^{+3.5}_{-0.9}$ & $9.$&$2^{+70}_{-2.8}$ &
      $1.1^{+1.3}_{-0.5}$ & $1.$&$1^{+0.7}_{-0.7}$ & $1.5^{+1.5}_{-0.5}$ &
      $8.$&$4^{+4.1}_{-2.8}$
      \\[0.65ex]
      \textsc{cemekl} & $0.95$ & $1.6^{+0.8}_{-0.7}$ & $43.7^{+3.2}_{-3.1}$ &
      $3.3^{+2.3}_{-1.3}$ & $0.72^{+0.09}_{-0.12}$ & $-$ &
      $11.$&$8^{+13.3}_{-5.1}$ & $1.2^{+1.3}_{-0.7}$ & $1.$&$6^{+2.1}_{-0.9}$ &
      $1.9^{+0.5}_{-0.5}$ & $6.$&$8^{+3.3}_{-1.9}$
      \\[0.65ex]
      \textsc{mkcflow} & $0.95$ & $1.4^{+0.4}_{-0.3}$ & $44.5^{+2.0}_{-0.5}$ &
      $4.0^{+1.0}_{-0.8}$ & $0.68^{+0.03}_{-0.06}$ & $<0.003$ &
      $48.$&$5^{+17.4}_{-4.9}$ & $1.7^{+1.1}_{-0.8}$ & $1.$&$3^{+0.4}_{-0.5}$ &
      $2.0^{+1.5}_{-0.8}$ & $6.$&$3^{+0.2}_{-0.2}$
      \\[0.65ex]
      accretion flow & $0.94$ & $1.5^{+0.8}_{-0.7}$ & $43.9^{+3.3}_{-3.2}$ &
      $3.3^{+2.5}_{-1.2}$ & $0.72^{+0.09}_{-0.12}$ & $0.8^{+0.4}_{-0.2}$ &
      $20.$&$0^{+9.9}_{-6.1}$ & $1.3^{+1.0}_{-0.6}$ & $1.$&$6^{+2.1}_{-0.8}$ &
      $2.0^{+1.5}_{-0.8}$ & $6.$&$6^{+3.2}_{-1.8}$ \\[0.65ex]

      \multicolumn{2}{l}{\emph{Bright phase 1}} & & & & & & & & & & \\
      \textsc{mekal} & $0.95$  & $1.9^{+1.0}_{-0.9}$ & $42.3^{+3.7}_{-3.5}$ &
      $1.8^{+2.2}_{-0.7}$ & $0.73^{+0.10}_{-0.15}$ & $-$ &
      $4.$&$6^{+4.5}_{-1.3}$ & $:=1.0$ & $4.$&$9^{+8.7}_{-3.0}$ &
      $5.6^{+5.5}_{-2.5}$ & $16.$&$4^{+11.9}_{-6.1}$
      \\[0.65ex]

      \multicolumn{2}{l}{\emph{Bright phase 2}} & & & & & & & & & & \\
      \textsc{mekal} & $1.03$ & $1.8^{+1.6}_{-1.3}$ & $44.0^{+6.1}_{-5.6}$ &
      $1.6^{+2.9}_{-0.8}$ & $0.82^{+0.11}_{-0.14}$ & $-$ &
      $4.$&$7^{+11.4}_{-2.2}$ & $:=1.0$ & $2.$&$0^{+7.9}_{-1.5}$ &
      $2.8^{+5.5}_{-1.6}$ & $8.$&$0^{+9.7}_{-3.7}$
      \\[0.65ex]

      \multicolumn{2}{l}{\emph{Faint phase}} & & & & & & & & & & \\
      \textsc{mekal} & $1.08$ & $3.2^{+4.3}_{-3.2}$ & $25.1^{+22.8}_{-19.2}$ &
      $3.4^{+3.5}_{-1.4}$ & $0.85^{+0.05}_{-0.07}$ & $-$ &
      $7.$&$1^{+4.8}_{-2.7}$ & $:=1.0$ & $2.$&$7^{+7.1}_{-7.1}$ &
      $2.0^{+1.1}_{-1.1}$ & $3.$&$4^{+8.7}_{-2.4}$
      \\[0.65ex]

      \multicolumn{2}{l}{\emph{Minimum phase}} & & & & & & & & & & \\
      \textsc{mekal} & $1.32$ & $< 3.0$ & $50.2^{+17.8}_{-12.6}$ &
      $14.7^{+11.6}_{-6.7}$ & $1.0$ & $-$ & $3.$&$2^{+9.9}_{-1.7}$ &
      $:=1.0$ & \multicolumn{2}{c}{$0.09^{+1.09}_{-0.03}$} &
      $0.15^{+0.06}_{-0.47}$ & $1.$&$1^{+1.1}_{-0.4}$
      \\[0.65ex]

      \multicolumn{2}{l}{\emph{2005/08/15}\tablefootmark{c}} & & & &
      & & & & & & \\
      \textsc{mekal} & $1.08$ & $3.6^{+0.7}_{-0.6}$ & $45.3^{+2.2}_{-2.2}$ &
      $15.6^{+15.5}_{-10.9}$ & $1.0$ & $-$ & $2.$&$7^{+16.0}_{-1.2}$ &
      $:=1.0$ & $25.$&$9^{+13.2}_{-7.7}$ & $38.0^{+13.4}_{-8.3}$ &
      \multicolumn{2}{c}{$104^{+42}_{-27}$}
      \\[0.2ex]
 
      \hline
    \end{tabular}
    }
    \tablefoottext{a}{The soft component in all models is a multi-temperature
      black body with maximum temperature $kT_\mathrm{bb}$.}
    \tablefoottext{b}{Bolometric black-body luminosities are determined as
      $L_\mathrm{bb,bol}=2\pi d^2F_\mathrm{bb}$, and the model fluxes by means
      of the \textsc{cflux} model.}
    \tablefoottext{c}{Simultaneous fit to EPIC and RGS data.}
  \end{table*}

%----------------------------------
\subsection{Soft X-ray component (accretion spot)}
\label{sec:softspec}

  The flux contribution of the white dwarf was described by absorbed black
  bodies, mainly forming the spectral range below 0.5\,keV and originating in
  the accretion-heated stellar surface.  The spectra exhibit deviations from a
  single black body that cannot be explained merely by uncertainties in the
  calibration. Within the accretion region, a wide range of temperatures is
  expected between the base of the accretion column and the unheated
  white-dwarf surface. To approximate this structure, we fitted
  multi-component black-body models, employing different Gaussian and
  exponential distributions of temperatures and emitting surface areas.  They
  improved the fit statistics in the energy range below 0.6\,keV to
  $\chi^2_\mathrm{red} = 1.1$, compared to $\chi^2_\mathrm{red} = 1.3$ for a
  single absorbed black body. For the final fit over the whole energy range,
  we used black bodies with a Gaussian distribution of temperatures and
  coupled emitting surface areas, fixing the minimum temperature at a typical
  white-dwarf surface temperature of $kT_\mathrm{bb,low} := 2\,\mathrm{eV}$
  ($\sim $23\,000\,K).  A maximum number of fifteen constituents were
  sufficient to reproduce a smooth temperature profile with small temperature
  steps, not exceeding 5\,eV between neighboring black-body components. The
  free parameters of the model were the temperature and the normalization of
  the hottest component. The model spectrum shows a shallower slope than a
  single black body. The highest X-ray flux arises from the hottest components
  with small effective areas, while the low-temperature black bodies covering
  larger surface areas contribute little to the total soft X-ray emission.
  The best-fit solution infers temperatures of as high as $kT_\mathrm{bb,high}
  = 43.9^{+3.3}_{-3.2}\,\mathrm{eV}$, which corresponds to a characteristic
  temperature $kT_\mathrm{bb}\sim 39.5\,\mathrm{eV}$ of a single black-body
  component. The hydrogen column absorption of $N_\mathrm{H,ISM} =
  1.5^{+0.8}_{-0.7}\times 10^{20}\,\mathrm{cm}^{-2}$ is on the same order as
  the Galactic $N_\mathrm{H}$ of up to $5\times 10^{20}\,\mathrm{cm}^{-2}$
  towards AI~Tri listed by \citet{kalberla:05} and \citet{dickey:90}.

%----------------------------------
\subsection{Hard X-ray component (post-shock accretion flow)}
\label{sec:plasma}

  To reproduce both the diffuse hot plasma and physical structure of the
  (post-)shock accretion column, we compared a number of different emission
  models (Table~\ref{tab:xspecfits}).  Fits involving several independent
  plasma components provide a smaller $\chi^2$ than those with one mean
  temperature and, thus, demonstrate that emission regions at different
  temperatures influence the spectral shape. Multi-temperature models in
  \textsc{xspec} comprise a sum of \textsc{mekal}\footnote{\textsc{mekal}
    \citep[e.\,g.][]{mewe:85,liedahl:95} models include spectral continuum and
    element lines from a hot, optically thin plasma based on the assumptions
    of collisional ionization equilibrium and Maxwellian electron and ion
    energy distributions. Calculating the continuum radiation involves
    free-free, free-bound, and two-photon emission.  Electron impact,
    radiative and dielectronic recombination, and inner-shell excitation and
    ionization contribute to the line transitions.}  emission model
  components. We tested \textsc{cemekl}, where the emission measures are
  proportional to a power law over temperature, and the cooling flow
  \textsc{mkcflow} (Table~\ref{tab:xspecfits}).  Using plasma models without
  considering the accretion physics, however, may lead to uncertain or
  erroneous bolometric flux estimates. Flux and luminosity ratios, used for
  investigating the energy balance (see Sect.~\ref{sec:fluxratio}), depend
  strongly on the model choice. We employ multi-temperature plasma models
  based on the emission region models of \citet{fischer:01}. They calculate
  the radiative transfer for bremsstrahlung and cyclotron radiation in a
  stationary two-fluid plasma. The local mass flow density is treated as the
  main free parameter and determines temperatures and shock height. For a
  characteristic primary mass of $M_\mathrm{WD} = 0.6\,M_{\sun}$, they derive
  temperature and density structures. We adopt these to account for the wide
  parameter range in the post-shock accretion column and to establish
  relations between the temperatures and the emission measures of several
  \textsc{mekal} components, assuming that optically thin conditions
  dominate. The element abundance, the emission measure, and the temperature
  of the coolest component are the free parameters for each $B-\dot{m}$
  combination. In addition, a common absorption term is needed for the total
  plasma spectrum, which accounts for intrinsic absorption by material
  surrounding the emission region in the accretion column.  We employ the
  partial-covering \textsc{pcfabs} and the solar metal abundance ratios of
  \citet{grevesse:07} in the plasma models.

\begin{figure}
    \centering
    \includegraphics[width=8.8cm]{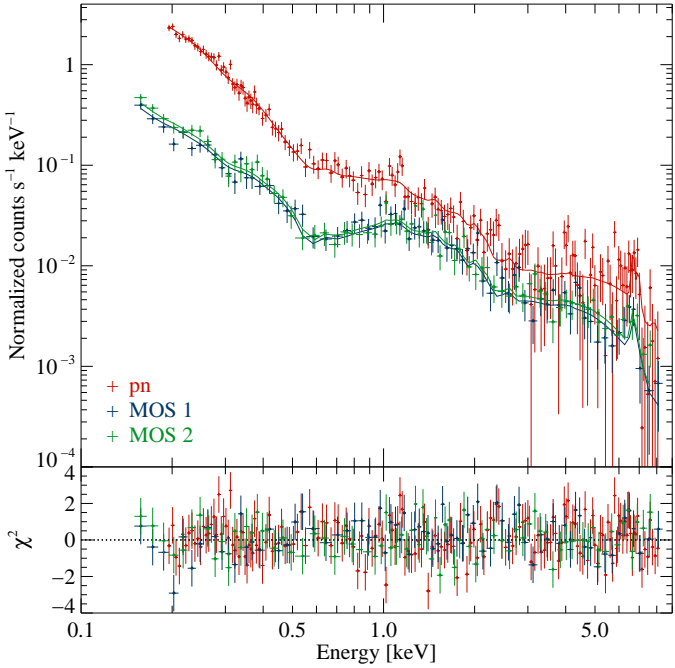}
    \caption{\footnotesize{EPIC spectra extracted from the 20\,ksec exposure
        of AI~Tri on August 22, 2005. The best-fit model consists of mildly
        absorbed black bodies, combined with partially absorbed
        multi-\textsc{mekal} plasma emission.}}
    \label{fig:spec1044}%
  \end{figure}

  In a single-temperature \textsc{mekal} fit to the accretion-column spectrum
  of AI~Tri, larger residuals remain around 1\,keV and between 3 and
  5\,keV. The multi-temperature plasma models describe the emission lines more
  precisely and improve the fit accuracy. The \textsc{cemekl} model clearly
  underestimates the continuum flux at energies above 7\,keV, while
  \textsc{mkcflow} and accretion-flow models yield similar parameters.
  Closest approximation of the spectral lines is achieved when using the
  multi-\textsc{mekal} accretion-flow spectrum with the temperature structure
  following the models of \citet[][``accretion flow'' model in
    Table~\ref{tab:xspecfits}]{fischer:01}. The magnetic field strength was
  fixed at $B = 40\,\mathrm{MG}$, referring to the result of
  \citet{schwarz:98}. From fits with mass flow densities between $\dot{m} =
  0.01$ and $100\,\mathrm{g\,cm^{-2}\,s^{-1}}$, an average $\dot{m} =
  0.1\,\mathrm{g\,cm}^{-2}\mathrm{s}^{-1}$ is derived. The best-fit solution
  infers a reduced $\chi^2_\mathrm{red} = 0.94$ at 318 degrees of freedom,
  plasma temperatures between $kT_{\textsc{mekal},\mathrm{low}}$ $=$
  $0.8^{+0.4}_{-0.2}\,\mathrm{keV}$ and $kT_{\textsc{mekal},\mathrm{high}}$
  $=$ $20.0^{+9.9}_{-6.1}\,\mathrm{keV}$, and $1.3^{+1.0}_{-0.6}$ times the
  solar abundances (Fig.~\ref{fig:spec1044}).

  Neither of the tested plasma models completely reproduces the spectral slope
  at energies above 4\,keV.  The deviations may be caused by remaining
  unresolved emission lines, which require a modified temperature structure or
  metal abundances. Alternatively, reflection of the continuum from the
  white-dwarf surface may harden the spectrum for energies above several keV,
  as \citet{done:95} expound for the GINGA spectra of \object{EF~Eri}. The
  residuals around 1\,keV can indicate photoionization effects. An additional
  warm absorber model component such as \textsc{absori}
  \citep{done:92,zdziarski:95} or \textsc{xstar} \citep{kallman:01,garcia:05}
  accounts more accurately for the spectral shape, but introduces too many
  free parameters.

%% EPIC spectra

%----------------------------------
\subsection{Phase-resolved modeling}
\label{sec:phaseres}

  The 20\,ksec XMM-Newton exposure was divided into four phase intervals,
  characterized by the varying amount of soft X-ray emission and the
  associated hardness ratios (cf.~Fig.~\ref{fig:lc1044}).  We extracted and
  modeled spectra for each of the intervals separately. Since the number of
  counts are lower than in the complete data set, we used absorbed multi-black
  body plus single \textsc{mekal} models. The element abundances were fixed to
  those of the solar distribution, which is approximately the orbital mean.

  During the two X-ray bright phases, the spectral parameters agree remarkably
  well. Both the black-body luminosity and emitting surface area exceed the
  values derived from the fits to the total spectrum, while the contribution
  of the plasma component remains nearly constant.  Owing to the higher
  background flux during the second bright phase, the fit is affected by
  larger uncertainties in the mean plasma temperature.  On the whole, the two
  bright phase spectra are identical within the error bars.
  
  Apart from an intermittent soft flare before $\varphi_\mathrm{phot} = 1.0$,
  the soft X-ray count rate is consistent with zero during the faint phase.
  The hardness ratio reaches a mean value of HR$_\mathrm{faint}=+0.53\pm0.01$
  and a maximum of one.  The model temperatures and the intrinsic absorption
  overlap with the bright-phase values within the large error bars and may be
  considered to be constant.  A low signal-to-noise caused by the high
  background count rate is the main source of the uncertainties in the fit.
 
  During the soft minimum phase interval, the low count rate also leads to a
  poorer fit accuracy.  The black-body flux diminishes, and the hardness ratio
  HR$_\mathrm{softmin}=-0.43\pm0.01$ is similar to the orbital mean of
  HR$_\mathrm{mean}=-0.43\pm0.09$.  In parallel, the intrinsic absorption
  increases substantially to $N_\mathrm{H,int} = 1.5^{+1.2}_{-0.7}\times
  10^{24}\,\mathrm{cm}^{-2}$, and the \textsc{mekal} norm rises, such that the
  plasma component contributes more significantly to the total X-ray flux
  within the XMM-Newton energy range.

%% Spectra 1041
  \begin{figure}
    \centering
    \includegraphics[width=8.8cm]{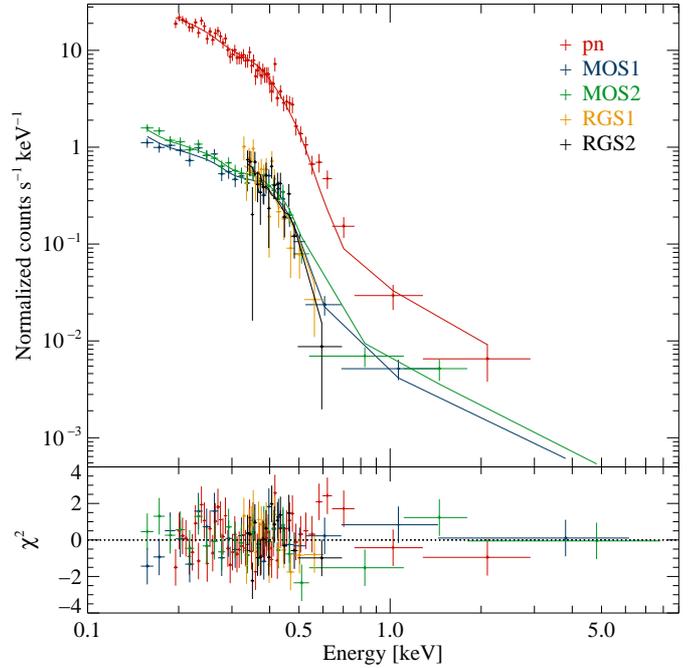}
    \caption{\footnotesize{Simultaneous fit to the XMM-Newton EPIC and RGS
        spectra on August 15, 2005 with the multi-temperature black-body plus
        single absorbed \textsc{mekal} model, as described in the text.}}
    \label{fig:spec1041}%
  \end{figure}

%----------------------------------
\subsection{XMM-Newton data on August 15, 2005}
\label{sec:aug15}

  From the 1.1 to 5\,ksec XMM-Newton observation on August 15, 2005, EPIC and
  RGS first order spectra were extracted and fitted simultaneously
  (Fig.~\ref{fig:spec1041}). Since most of the signal was detected at energies
  below 1\,keV (HR$=-0.94\pm 0.06$) and the hard X-ray component almost
  vanished, we chose an absorbed multi-temperature black-body model plus a
  single \textsc{mekal} with fixed solar element abundance.  The black-body
  parameters differ only marginally from the best fit to the 20\,ksec exposure
  on August 22, 2005. The source flux estimated from the \textsc{xspec} fits,
  however, was at least a factor of ten higher than that during the 20\,ksec
  exposure one week later. The significant increase in the flux was caused
  mainly by the dominant soft X-ray component and a larger emitting surface
  area.

%% EPIC & OM
  \begin{figure}
    \centering
    \includegraphics[width=8.8cm]{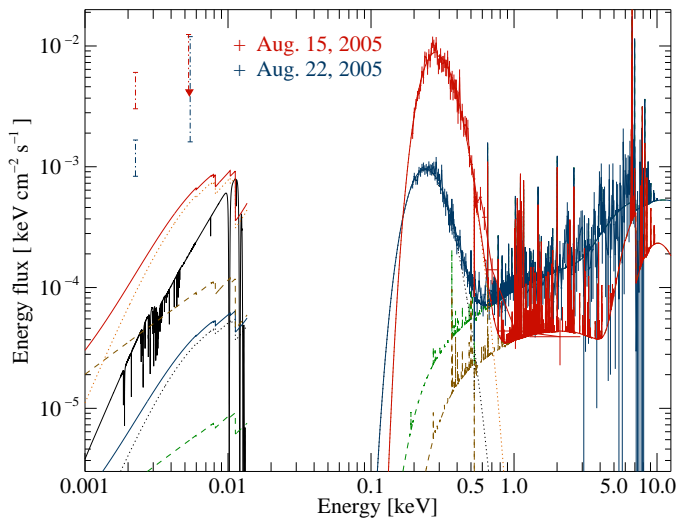}
    \caption{\footnotesize{Spectral energy distribution of AI~Tri. To the
        right, EPIC/pn spectra from both the XMM-Newton pointings on August 15
        and 22, 2005. To the left, the corresponding OM/UVM2 spectral points
        and the $V\!$-band observations on August 17 and 29, respectively,
        each with orbital minimum and maximum flux.  On August 15, the UV
        minimum flux is unknown for lack of phase coverage. Overplotted are
        the best-fit models (solid lines) as described in the text, their
        individual components absorbed black body (dotted) and \textsc{mekal}
        (dashed) for each observation, and a non-LTE white-dwarf model
        spectrum at $T_\mathrm{eff}=20\,000~\mathrm{K}$ (black).}}
    \label{fig:spec1044all}%
  \end{figure}

%----------------------------------
\subsection{UV flux}
\label{sec:uvflux}  

  To extend the spectral energy distribution coverage, we converted
  photometric measurements obtained at ultraviolet and optical wavelengths
  into energy flux units. The total OM/UVM2 flux in the range between
  2050\,{\AA} and 2450\,{\AA} was calculated from the corrected count rate
  using the conversion factor from white dwarf standards, where
  1~count\,s$^{-1}$ is equivalent to
  $2.2\times10^{-15}\ \mathrm{erg~cm}^{-2}\,\mathrm{s}^{-1}${\AA}$^{-1}$. It
  is similar in both the observations. Figure~\ref{fig:spec1044all} shows the
  spectral points from our OM/UVM2 and $V\!$-band photometry, the EPIC/pn
  spectra, and the best-fit black-body plus \textsc{mekal} models. To be able
  to compare with the flux contribution of the unheated primary, a
  hydrogen-helium non-LTE white-dwarf atmosphere spectrum was calculated using
  the modeling package NGRT \citep{werner:99}, whose predicted flux can serve
  as an approximate lower limit to the low-energy emission of the system. We
  assume a distance of $d\sim620\,\mathrm{pc}$ \citep{schwarz:98} and a
  typical effective temperature of $T_\mathrm{eff}=20\,000~\mathrm{K}$ for the
  white dwarf (Fig.~\ref{fig:spec1044all}).  The orbital means of optical and
  UV flux exceed the models by factors up to 40 at both observational
  epochs. This is not surprising, as other system components such as the
  accretion stream can contribute significantly to the optical and UV flux. It
  is also consistent with the optical spectra presented by \citet{schwarz:98},
  in which no signatures of the white dwarf could be detected.  The emitting
  surface area of the low-temperature model components may be, in addition,
  significantly larger than derived from the X-ray spectra. The UV data points
  can thus be taken to represent an upper limit, which is not violated by the
  spectral components required to model the X-ray range.

%
%---------- DISCUSSION ----------

%
%---------- Soft-to-hard ratio ----------
%
\section{On the soft-to-hard X-ray flux ratios}
\label{sec:fluxratio}

  Several polars have a distinctly higher luminosity at soft than at hard
  energies. The true fraction of systems exhibiting this 'soft X-ray excess'
  continues to be a topic of discussion. The phenomenon is apparently
  connected to the scenario of inhomogeneous and time-dependent (`blobby')
  column accretion proposed by \citet{kuijpers:82} and developed further by
  \citet{frank:88}. The amount of the excess increases with magnetic field
  strength \citep{beuermann:94,ramsay:94}. \citet{beuermann:95} point out that
  at a high field strength ($B>30\ \mathrm{MG}$) and a relatively low
  accretion rate, cyclotron emission rather than bremsstrahlung becomes
  the main cooling process, causing an enhanced soft X-ray flux. Mass flow
  rate, magnetic field strength, and white-dwarf mass thus regulate the energy
  balance in polars \citep[cf.][]{fischer:01}.  \citet{ramsay:04balance}
  examine new XMM-Newton and re-calibrated ROSAT data. Considering geometrical
  effects and the influence of accretion blobs, they derive soft-to-hard
  ratios close to one for the majority of the observed systems and conclude
  that fewer polars than previously understood exhibit a notable soft X-ray
  excess.  With its strong magnetic field and known high soft X-ray emission,
  AI~Tri is a suitable target to investigate the potential connection between
  system parameters, flaring signals from blobby accretion, and flux ratio.

%----------------------------------
\subsection{Photometric signs of blobby accretion}
\label{sec:fluxratiophot}

  Rapid variations characterize the optical as well as the X-ray light curves
  and are most likely caused by inhomogeneous accretion.  Prominent flares may
  occur, when dense blobs impinge on the white-dwarf atmosphere. In the X-ray
  light curves, individual accretion events cannot be identified, though their
  shapes are probably formed by superimposed emission from multiple blobby
  events.  This differs from the situation for the purely blobby accreting
  long-period polar \object{V1309~Ori}, where \citet{schwarz:05} distinguished
  single accretion flares. The possible quasi-periodic oscillations in the
  optical light curves (Sect.~\ref{sec:shortperiod}), which may arise from
  blob-excited oscillations of the shock front, are in addition indicative of
  inhomogeneous accretion. \citet{shafter:95} also report on QPOs at multiple
  frequencies in optical light curves of V1309~Ori.

%----------------------------------
\subsection{Flux ratios from multi-temperature spectral modeling}
\label{sec:fluxratiospec}

  The soft-to-hard X-ray flux ratios can be derived from unabsorbed bolometric
  luminosities.  The observed data, however, cover only part of the spectral
  energy distribution, so the total flux has to be extrapolated and thus
  strongly depends on the choice of the model and its assumptions. In
  particular, results from former ROSAT analyses are uncertain, since the hard
  X-ray component had to be approximated with bremsstrahlung and the
  temperature could not be constrained from the spectral fits.

  Our multi-temperature black-body and \textsc{mekal} model fit to the 2005
  August 22 XMM-Newton spectra yields the typical parameters of the
  high-energy distribution of CVs. Using the distance estimate of AI~Tri
  \citep{schwarz:98} $d=620\pm 100\,\mathrm{pc}$, it results in an unabsorbed
  black-body luminosity of $L_\mathrm{bb} = 2.0^{+1.5}_{-0.8}\times10^{32}
  (d/620\mathrm{pc})^2\,\mathrm{erg\,s}^{-1}$ and an integrated X-ray flux
  $F_\mathrm{bol} = 1.3^{+0.7}_{-0.3}\times
  10^{-11}\,\mathrm{erg\,cm}^{-2}\,\mathrm{s}^{-1}$.  The flux in the soft
  $0.1-0.5\,\mathrm{keV}$ energy band exceeds the hard
  $0.5-10.0\,\mathrm{keV}$ flux during the whole orbital revolution. The
  bolometric fluxes of the individual unabsorbed model components lead to
  ratios of $F_\mathrm{bb}/F_\textsc{mekal} \ga 2.1^{+1.6}_{-0.8}$ for the
  orbital mean and of $({F_\mathrm{bb}}/{F_\textsc{mekal}})_\mathrm{bright 1}
  \ga 5.7^{+5.6}_{-2.5}$ for the first bright phase.  Owing to the
  partial-covering absorption term employed in the spectral fits, the plasma
  temperatures are lower than in models that do not account for intrinsic
  absorption. In particular, the spectral resolution and range of the ROSAT
  spectra of polars was not high enough to derive both a reliable
  bremsstrahlung temperature and the absorbing column density.  Hence, the
  cooler plasma component contributes more significantly to the emission in
  the soft X-ray regime, resulting in a lower ratio of the extrapolated fluxes
  of the black-body to the \textsc{mekal} components.

  On August 15, 2005, the total integrated flux $F_\mathrm{bol} =
  1.6^{+0.8}_{-0.5}\times10^{-10}\,\mathrm{erg\,cm}^{-2}\,\mathrm{s}^{-1}$ was
  notably governed by an increased black-body component, while the model flux
  in the hard band stays at a low level. Due to the low count rate above
  energies of 0.8\,keV, extrapolating the model flux to a wider energy range
  would have produced unacceptably large uncertainties. We therefore used the
  flux ratio in the XMM-Newton bands, $F_\mathrm{soft}/F_\mathrm{hard} =
  130^{+95}_{-60}$, as a lower limit to the bolometric flux ratio.  The
  black-body flux is remarkably higher than on August 22, indicating that the
  mass accretion rate is higher by a factor of more than one hundred. This may
  induce a stronger temporal and spatial inhomogeneity in the accretion
  stream, forcing more material into dense blobs. In consequence, the
  contribution of the hard X-ray emission will be lower than in states of
  lower accretion rates, producing a distinct change in the spectral shape as
  seen in Fig.~\ref{fig:spec1044all}. The pronounced flickering in the X-ray
  light curves supports the idea of blobby accretion playing a larger role at
  that time.

  An extreme dominance of the soft X-ray component was also found in the
  XMM-Newton spectra of V1309~Ori with a bolometric flux ratio of several
  thousand \citep{schwarz:05} and of the short-period polar \object{EU~UMa}
  with a luminosity ratio of $14-2800$ \citep{ramsay:04shortper}. In the few
  systems lacking distinct soft X-ray emission, such as
  \object{2XMMp\,J131223.4+173659} \citep{vogel:08} and \object{V2301~Oph}
  \citep{ramsay:07}, the flux maximum of the reprocessed component appears to
  be shifted towards the EUV and UV regime.  The same could apply to objects
  that show a detectable, but very weak soft component, such as
  \object{EV~UMa} \citep{ramsay:03}. This interpretation implies that single
  black-body models, adjusted to the X-ray regime, generally give only a lower
  limit for the reprocessed emission from the accretion region and thus for
  the flux ratio. To quantify the soft X-ray excess, multi-temperature models
  are required not only to describe the emission from the post-shock accretion
  column but also to adequately take into account the soft X-ray emission from
  the accretion area on the white dwarf.

%
%---------- Geometry ----------
%
\section{Accretion geometry of the system}
\label{sec:geometry}

  Depending on the field topology, the magnetic flux density, and the ram
  pressure of the free-falling accretion stream, coupling of the accretion
  plasma onto the field line trajectories can occur at different locations
  within the binary reference frame, and the infalling material may be
  channeled into one or more accretion regions on the white dwarf
  surface. Constraining the geometry and location of the active X-ray emitting
  accretion region(s) is important to understanding the origin of X-ray soft
  and hard components and their flux balance.

  Prominent light curve features and phase-resolved spectral modeling may
  provide insight into the system geometry. Since it is crucial to know their
  orbital phasing, we convert the photometric into orbital phases as described
  in Sect.~\ref{sec:optphot} and refer to these values in the following
  discussion, which is valid based on the assumption of synchronous rotation.
  These features are for instance the two bright phases in our X-ray light
  curves and the minimum in-between. The almost identical fits to the
  bright-phase spectra strongly indicate that the complete soft X-ray emission
  originates in one and the same accretion region. The minima around
  $\varphi_\mathrm{orb} = 0.26$ and $0.81$ ($\varphi_\mathrm{phot} = 0.45$ and
  $0.0$) are pronounced mainly in the soft X-ray flux and may have their
  origin in a variety of mechanisms. These include a total or partial
  self-eclipse of the accretion region by the white dwarf, absorption in the
  accretion stream, or random mass-transfer variations on timescales of
  several hours.  An occultation of the white dwarf by the secondary star can
  be excluded, since it would be expected to occur at inferior conjunction.

%----------------------------------
\subsection{Self-eclipse of the accretion region}
\label{sec:selfeclipse}

  A self-eclipse occurs when the accretion region passes behind the limb of
  the white dwarf because of the stellar rotation. Such a feature was found in
  the X-ray / EUV light curves of several polars such as \object{AR~UMa}
  \citep{szkody:99}, MT~Dra \citep{schwarz:02}, VV~Pup \citep{vennes:95}. The
  X-ray spectra during phases of self-eclipse are characterized by a vanishing
  emitting area, at least of the soft X-ray component, while the intrinsic
  absorption remains nearly constant.

  At a system inclination of $i\sim 70\degr$ \citep{katajainen:01}, a partial
  or total self-eclipse is plausible for AI~Tri, if the colatitude of the
  accretion region is higher than $20\degr$.  Of the two phase intervals with
  reduced X-ray emission detected for AI~Tri, the major faint phase exhibits
  the characteristic signatures of a self-eclipse, indicating that larger
  parts of the accretion region might disappear behind the horizon of the
  white dwarf between $\varphi_\mathrm{orb} = 0.61$ and $\varphi_\mathrm{orb}
  = 0.86$. During the $\varphi_\mathrm{orb} = 0.21 - 0.31$ interval, the
  hardness ratio is not higher than zero, and soft and hard X-radiation are
  emitted at a similar level. An eclipse of the accretion region, hence, is
  less likely to explain this phase.

%----------------------------------
\subsection{Stream absorption}
\label{sec:streamabs}

  When the accretion column crosses the line of sight, it can obscure parts of
  the accretion region. This manifests itself as a sharp dip in the soft X-ray
  light curves with a typical duration of $\Delta\varphi \sim 0.1$ and as
  significantly enhanced absorption in the X-ray spectrum. Light curves and
  spectra of AI~Tri display this behavior during the soft X-ray minimum around
  $\varphi_\mathrm{orb} = 0.26$. The partial-covering absorption term in the
  spectral models rises markedly; the hardness ratio of the light curves
  increases from $\mathrm{HR} \sim -0.8$ to values around zero
  (Fig.~\ref{fig:lc1044}). The short ingress and egress times and the duration
  $\Delta\varphi = 0.1$ of the light curve dip are also consistent with an
  origin in stream absorption.  The energy-dependent dip due to stream
  absorption and the broad faint phase in the X-ray light curves of AI~Tri
  resemble the properties of the high-field system AR~UMa \citep{szkody:99}
  and the two-pole accretor MT~Dra \citep{schwarz:02}. Their X-ray emission
  can be explained in the same way.

%% Sketch of binary geometry
  \begin{figure}
    \centering
    \includegraphics[width=8.cm]{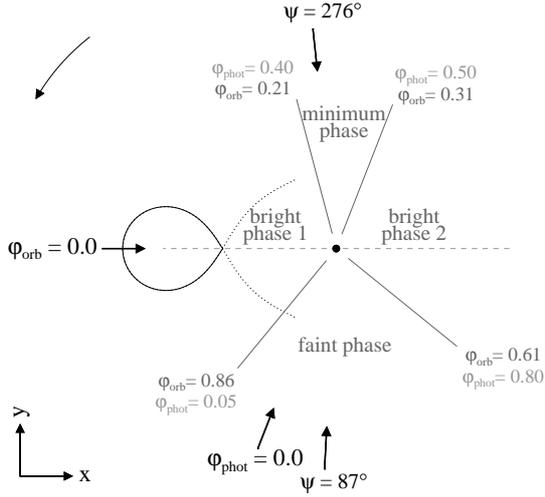}
    \caption{\footnotesize{Sketch of the binary geometry on the basis of the
        spectroscopic phasing derived by \citet{schwarz:98}, valid in the case
        of synchronous rotation, the Roche lobes being calculated for a mass
        ratio of $Q = \mathrm{M}_\mathrm{WD}:\mathrm{M}_2 = 1.5$. The
        individual phase intervals are marked to illustrate the formation of
        the most prominent light curve features.}}
    \label{fig:sketch}
  \end{figure}

%----------------------------------
\subsection{A second possible scenario}
\label{sec:altint}

  The faint phase might alternatively be interpreted as stream absorption and
  the X-ray light curve dip at $\varphi_\mathrm{orb} = 0.26$ as a self-eclipse
  of the accretion region.  In consequence, the optical and the UV light
  curves would then be anti-correlated with the X-ray data, which would
  disagree with the standard picture of accretion in polars
  \citep[e.\,g.][]{king:79,lamb:79} in which the maximum flux in the UV range
  is seen when the accretion region passes the line of sight -- just at the
  phase of the stream eclipse. If the ultraviolet light curve, however, were
  dominated by the revolution of an extended accretion column, we may see its
  irradiated surface in anti-phase with the soft X-ray
  emission. \citet{osborne:87} describe an anti-correlation of optical and
  soft X-ray flux for \object{QQ~Vul}, another long-period polar, and consider
  a possible connection to accretion by means of dense filaments.

  However, the length of the faint phase, the slow ingress, and the low
  intrinsic absorption do not agree well with the characteristics of a stream
  dip. The short soft minimum around $\varphi_\mathrm{orb} = 0.26$ cannot be
  identified in the former ROSAT light curves. A phase of self-eclipse,
  however, would be expected to recur on longer timescales. For both reasons,
  we consider this second scenario less likely.

%----------------------------------
\subsection{Orientation of the main accretion pole}
\label{sec:orient}

  According to the standard picture of the accretion geometry in polars,
  described for example by \citet{cropper:88}, the magnetic pole of the white
  dwarf onto which the major part of the accretion stream is directed is
  normally found between longitudes of $\psi = 0\degr$ and $90\degr$. In the
  alternative interpretation of our X-ray light curves presented in
  Sect.~\ref{sec:altint}, the accretion region of AI~Tri would be visible from
  $\varphi_\mathrm{orb} = 0.31$ until $\varphi_\mathrm{orb} = 1.21$ and, thus,
  be located around $\psi = 87\degr$. In our favorite scenario
  (Sects.~\ref{sec:selfeclipse} and \ref{sec:streamabs}), on the other hand,
  the bright-phase emission is centered around $\varphi_\mathrm{orb} \sim
  0.23$, which corresponds to an unusual longitude $\psi = 280\degr$ of the
  dominating pole. But a standard geometry can also be preserved in this case,
  if the system rotates asynchronously. Asynchronous rotation cannot be
  excluded, as most of our observations (including the ROSAT and XMM-Newton
  data) trace solely the spin period. The information about both the orbital
  period and the phase shift between photometric and spectroscopic ephemeris
  is derived from the optical spectroscopy of \cite{schwarz:98} in 1992/93,
  who report an upper limit on the order of $10^{-4}$ for the degree of
  synchronization.  To place the accretion region of AI~Tri within the first
  quadrant, the orbital phasing at the epoch of the XMM-Newton observations
  had to be shifted by $\Delta\varphi_\mathrm{orb} \sim 0.2-0.5$.  A phase
  shift of this order can be achieved by a small degree of asynchronism of at
  least $|P_\mathrm{orb}-P_\mathrm{spin}|/P_\mathrm{orb}$ $=$
  $2-5\times10^{-5}$. This is even below the value of $0.28\,\%$ that
  \citet{staubert:03} derive for \object{V1432~Aql}, the polar with the
  slightest non-synchronous rotation determined up to now. \citet{campbell:99}
  indeed suggest that a significant number of polars with an undetected slight
  asynchronism may exist.

%----------------------------------
\subsection{The hard X-ray emission}
\label{sec:hard}

  Low hard X-ray flux remains detectable throughout the orbital cycle. The
  emission may originate in an extended accretion region that is not totally
  eclipsed. Another possible explanation is a second accretion pole that is
  only visible in the high energy range and accretes at a low mass-flow rate.
  Episodes of accretion onto two poles with different hardness ratios were
  identified for example in AM~Her \citep[e.\,g.][]{matt:00}, in
  \object{MT~Dra} \citep{schwarz:02}, and in \object{BY~Cam}
  \citep{ramsay:02}, one of the few known asynchronous systems.

%----------------------------------
\subsection{Long-term variation}
\label{sec:longterm}

  The X-ray light curves obtained with ROSAT and XMM-Newton between January
  1991 \citep{schwarz:98} and August 2005 (this work) do not completely repeat
  themselves.  A pronounced faint phase before or around
  $\varphi_\mathrm{phot} = 0.0$ ($\varphi_\mathrm{orb} = 0.8$), though,
  appears to be common to all X-ray light curves, which is consistent with the
  characteristics of a self-eclipse. It largely coincides with the UV and
  optical minima and with a peak in the linear polarization
  \citep{katajainen:01}.  The ROSAT HRI light curves taken between January 15
  and February 5, 1998 are similar to the XMM-Newton light curves, but lack
  the deep soft X-ray minimum around $\varphi_\mathrm{phot} = 0.45$
  ($\varphi_\mathrm{orb} = 0.26$), identifiable in the XMM-Newton data on
  August 22, 2005.

% Accretion curtain
  The accretion scenario described above cannot explain all the phenomena in
  the XMM-Newton and archival ROSAT data. In particular, the emission during
  the second bright phase is lower than during the first one; and the count
  rate changes distinctly between the different X-ray observations. An
  expanded accretion curtain along the ballistic accretion stream might reduce
  the flux between $\varphi_\mathrm{orb} = 0.3$ and $0.7$, as confirmed in the
  case of \object{HU~Aqr} by \citet{schwope:01}. Such an accretion curtain
  would account for the slightly enhanced absorption inferred by the spectral
  fits to the XMM-Newton data during the second bright phase, and be
  consistent with the finding that the narrow emission line components in the
  optical spectra are asymmetric \citep{schwarz:98}. By more or less effective
  shielding, the accretion curtain could induce the other strong variations
  detected in the X-ray light curves. \citet{schwope:01} report similar
  effects in the light curves and spectra of HU~Aqr.

% Changes in geometry
  Other possible reasons for the long-term evolution of the light curves are
  switches between one-pole and two-pole accretion or a rearrangement of the
  accretion geometry. The location of the threading region, where the ionized
  matter in the accretion stream attaches itself to the magnetic field lines
  and is lifted off the orbital plane, depends on the mass accretion rate
  $\dot{M}$. A change in $\dot{M}$ will influence the stream trajectory and
  thus the region of impact onto the white dwarf at the footpoints of the
  magnetic field lines, whose spread determines the form, extent, and location
  of the accretion region. A large number of dense filaments in the stream in
  addition affects its structure. The phase shift of the irregular optical
  light curve minimum on August 17, 2005 could reflect this modified accretion
  geometry. The spectral fits to the 2005 August 15 XMM-Newton data indicate a
  highly increased mass accretion rate at this epoch
  (Sect.~\ref{sec:fluxratiospec}). Inhomogeneities and changes in the emission
  region, hence, are likely. The irregular mode of the $V\!$-band light curves
  on October\,/\,November 1992 \citep{schwarz:98} could be explained in the
  same way.

%
%---------- Conclusions ----------
%
\section{Conclusions}

  By analyzing our phase-resolved X-ray spectroscopy with XMM-Newton, we have
  confirmed a strong soft X-ray excess of AI~Tri. Both the emission from the
  heated accretion spot and the emission from the post-shock accretion flow
  depart from single temperature models. From our multi-temperature
  approaches, we have found a bolometric flux ratio of
  ${F_\mathrm{bb}}/{F_\textsc{mekal}} \ga 5.7^{+5.6}_{-2.5}$ between the two
  components during the bright phase of AI~Tri on August 22, 2005. An even
  higher flux ratio of $F_\mathrm{soft}/F_\mathrm{hard} \ga 130^{+95}_{-60}$
  in the XMM-Newton energy bands was observed on August 15, 2005, when the
  total source flux was by a factor of at least ten higher. The distinct soft
  X-ray excess is probably related to inhomogeneous, blobby accretion as
  indicated by the high variability of the optical and X-ray flux on short
  timescales.

  From the light-curve characteristics and the spectral variation in the
  intrinsic absorption with phase, we have inferred the most likely geometry
  of AI~Tri. One main soft X-ray emitting accretion region undergoes a
  self-eclipse and a short period of accretion-stream absorption during the
  orbital revolution of the system. In this interpretation, the maxima of the
  UV and optical light curves occur at the moment that the accretion region
  passes the line of sight. The finding that hard X-ray emission is present at
  an almost constant, very low level over the whole orbital cycle, including
  the faint phase, may imply that a second accretion region with weak hard
  X-ray emission exists.  Changes in the accretion geometry may occasionally
  occur, possibly in response to significant variations in the mass accretion
  rate. Signatures of these changes are the irregular light curves on October/
  November 1992 and on August 17, 2005, and the new minimum dip seen in the
  $V\!$-band light curves from November 2006 onward.

  Assuming synchronous rotation of the white dwarf with the binary orbit, the
  main accretion region appears to be located around an unusual longitude of
  $\psi=275\degr$. Since we extrapolated the spectroscopic ephemeris from data
  of \citet{schwarz:98}, obtained more than ten years before our
  multiwavelength observations, we cannot exclude there being a slightly
  asynchronous rotation of AI~Tri within the accuracy reached. Future optical
  spectroscopy is necessary to clarify the ephemeris and to finally decide on
  the location of the accretion pole(s).

%
%---------- Acknowledgments ----------
%
\begin{acknowledgements}
  We thank F.~V.~Hessman and T.~Nagel for carrying out the $V\!$-band
  observations at the \textsc{Monet}/North and T\"ubingen telescopes, and the
  referee for constructive remarks. IT gratefully acknowledges helpful
  discussions with S.~Carpano, S.~Fritz, D.~deMartino, C.~Mauche, and
  J.~Wilms. This research is granted by DLR under project numbers 50\,OR\,0404
  and 50\,OR\,0501.

\end{acknowledgements}

%
%---------- References ----------
%
\bibliographystyle{aa}
\bibliography{13201}

\end{document}